\documentclass[journal=jpccck,manuscript=article]{achemso}
\usepackage[version=3]{mhchem} 
\usepackage[latin1]{inputenc}
\usepackage{scalefnt}
\usepackage{graphicx, subfigure}
\usepackage{xcolor}
\usepackage{braket}
\usepackage{amsmath}
\usepackage{float}
\usepackage{placeins}
\usepackage{afterpage}
\usepackage{units}

\usepackage[misc,geometry]{ifsym} 

\title[\texttt{achemso} demonstration]
{Switching the Conductance of a Molecular Junction using a Proton Transfer Reaction}
\author{Chriszandro Hofmeister}
\author{Rainer H\"{a}rtle}
\altaffiliation{Current address: Institut f{\"u}r Theoretische Physik, Friedrich-Hund-Platz 1, D-37077 G{\"o}ttingen, Germany}
\author{{\'O}scar Rubio-Pons}
\affiliation[Friedrich-Alexander Universit\"{a}t Erlangen-N\"{u}rnberg]
{Institut f\"{u}r Theoretische Physik und Interdisziplin\"{a}res Zentrum f\"{u}r Molekulare Materialien (ICMM), 
Friedrich-Alexander Universit\"{a}t Erlangen-N\"{u}rnberg, Staudtstra{\ss}e 7/B2 91058, Erlangen, Germany}
\author{Pedro B. Coto}
\email{pedro.brana-coto@physik.uni-erlangen.de}
\affiliation[Friedrich-Alexander Universit\"{a}t Erlangen-N\"{u}rnberg]
{Institut f\"{u}r Theoretische Physik und Interdisziplin\"{a}res Zentrum f\"{u}r Molekulare Materialien (ICMM), 
Friedrich-Alexander Universit\"{a}t Erlangen-N\"{u}rnberg, Staudtstra{\ss}e 7/B2 91058, Erlangen, Germany}
\author{Andrzej L. Sobolewski}
\email{sobola@ifpan.edu.pl}
\affiliation[Polish Academy of Sciences]
{Institute of Physics, Polish Academy of Sciences, PL-02668 Warsaw, Poland}
\author{Michael Thoss}
\email{michael.thoss@physik.uni-erlangen.de}
\affiliation[Friedrich-Alexander Universit\"{a}t Erlangen-N\"{u}rnberg]
{Institut f\"{u}r Theoretische Physik und Interdisziplin\"{a}res Zentrum f\"{u}r Molekulare Materialien (ICMM), 
Friedrich-Alexander Universit\"{a}t Erlangen-N\"{u}rnberg, Staudtstra{\ss}e 7/B2 91058, Erlangen, Germany}

\begin{document}

\maketitle

\begin{abstract}
A novel mechanism for switching a molecular junction based on a proton transfer
reaction triggered by an external electrostatic field is proposed. 
As a specific example to demonstrate the feasibility of the mechanism, 
the tautomers [2,5-(4-hydroxypyridine)] and 
\{2,5-[4(1H)-pyridone]\} are considered.
Employing a combination of first-principles electronic structure calculations
and Landauer transport theory, we show that both tautomers 
exhibit very different conductance properties and realize the
``on'' and ``off'' states of a molecular switch. Moreover, we
 provide a proof of principle that both 
forms can be reversibly converted into each other using an external 
electrostatic field.
\end{abstract}

\section{Introduction}
\label{intro}

Molecular junctions, where a single molecule 
is bound to two electrodes, provide interesting systems to study basic
mechanisms of non-equilibrium charge transport at the nanoscale  
and are promising candidates 
for the development of nanoelectronic devices \cite{Park2000,Elbing2005}. 
The  possibility of using single molecules as electronic components in 
electronic circuits has motivated intensive experimental 
\cite{Park2000,Elbing2005,Reed1997,Reichert2002,Smit2002,Qiu2004,Tao2006,Ioffe2008,Meded2009,vanderMolen2009,vanderMolden2010,Ballmann2012,Ballmann2013} 
and theoretical \cite{ChemPhys2002,Nitzan2003,Cuniberti2005,Joachim2005,Datta2005,Galperin2008,Nazarov2009,Cuevas2010,Hartle2011,Zimbovskaya2011,Thoss2013,Hartle2013,Pshe2013} research on the conductance properties of 
these systems and revealed a multitude of interesting transport phenomena.
In particular, it has 
been demonstrated that a molecular junction may be used as a 
nano switch if the molecule bridge has two 
(or more) stable states with different conductance, which can be reversibly 
transformed into each other \cite{vanderMolden2010,Aviram1974,Metzger2008,Vuillaume2008}.
 
In recent years,  
several optical and non-optical mechanisms for molecular switches in junctions
have been investigated 
 \cite{vanderMolden2010}. 
Many of the optical mechanisms 
rely on conformational changes of the molecule triggered by light 
absorption (such as light induced {\em cis}-{\em trans} 
isomerization processes \cite{Choi2006,Cuniberti2007}) to facilitate 
reversible switching between different conductance states. 
Non-optical 
mechanisms investigated include electrochemical switching, 
where a reversible redox 
pair is generated using voltage pulses \cite{vanderMolden2010,Mendes2005} 
or mechanical 
switching, where external mechanical forces are applied to produce
 the reversible change between states with different conductance
 \cite{Prasong2013}.

In recent theoretical work by some of the authors \cite{Sobo2008,Benesch2009}, 
it was shown that it is possible to use a photoinduced excited state 
hydrogen transfer reaction between the 
keto and enol tautomers of a molecule
to realize a molecular switch. As was demonstrated for an example of
 a polyene 
functionalized with keto and amino groups, this switching 
mechanism relies on the fact that keto and enol tautomeric forms often
exhibit different conjugation properties, which are associated with 
localized or delocalized charge distributions on the molecular bridge, 
and thus realize different conductance states.

\begin{figure}
\centering
\includegraphics[width=0.55\textwidth]{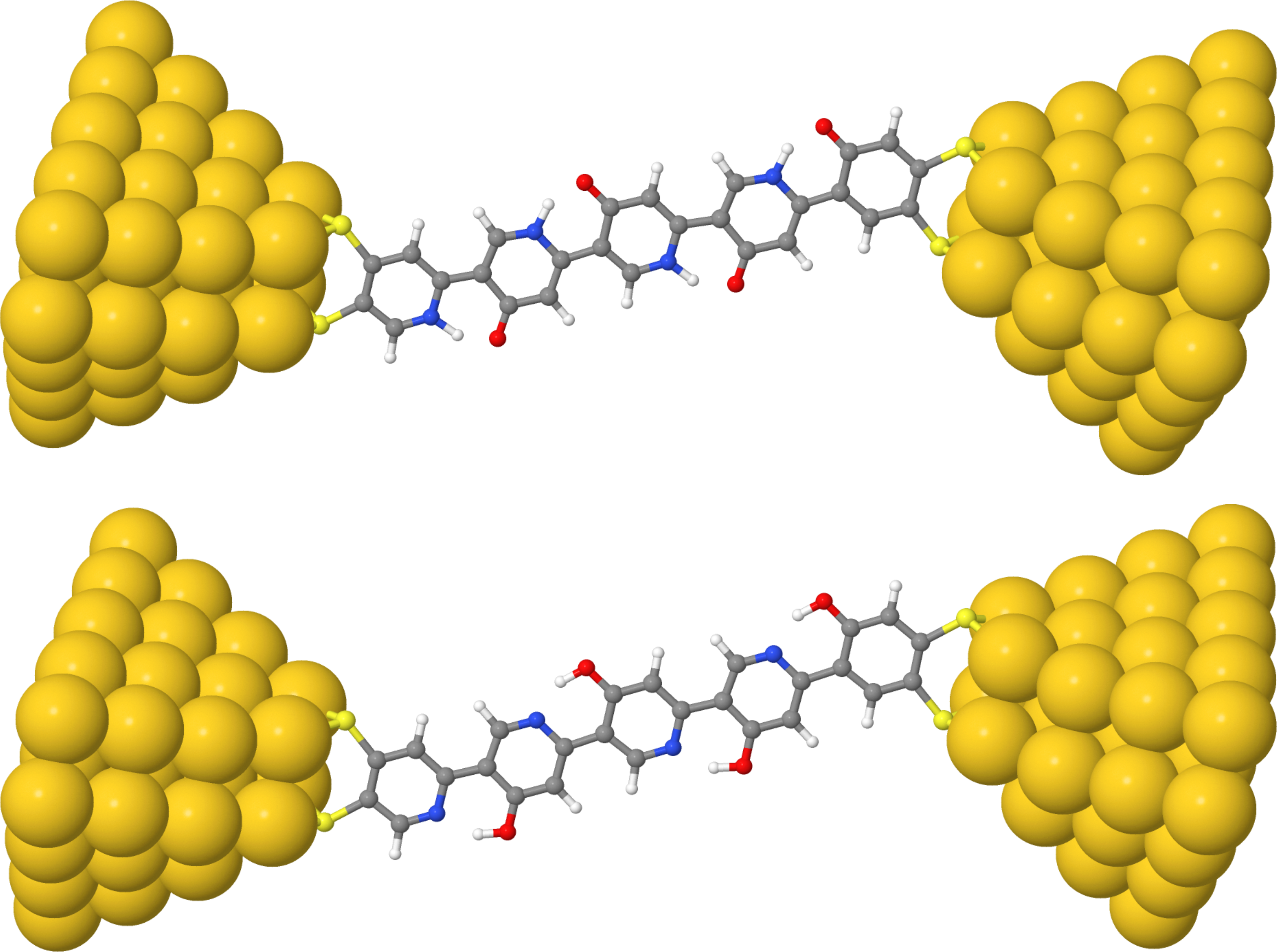}
\caption{\label{fig2} Keto tris\{2,5-[4(1H)-pyridone]\} (T2PY, top) and enol tris[2,5-(4-hydroxypyridine)] (T2HP, bottom) tautomers of the molecular 
switch investigated in this work. Both forms have been functionalized  for binding to the Au leads with 3,4-dimercapto-(1H)-pyridine and 3,4-dimercaptobenzenone (keto form, top) and 3,4-dimercaptopyridine and 3,4-dimercaptophenol (enol form, bottom).}     
\end{figure}

In this contribution, we build on these results and propose a 
new non-optical switching mechanism based on a ground state proton transfer 
reaction triggered by an external electrostatic field. 
As an example to demonstrate the feasibility of this mechanism, we 
study the ground state proton transfer reaction between oligomeric forms of 
p2HP and p2PY. Specifically, we investigate the proton transfer reaction 
between the enol tris[2,5-(4-hydroxypyridine)] (T2HP) and the corresponding 
keto form tris\{2,5-[4(1H)-pyridone]\} (T2PY) (for binding to the Au leads, 
the oligomers have been appropriately functionalized at their ends as indicated in Fig. \ref{fig2}). 
We show that both tautomers exhibit different current-voltage 
characteristics providing the ``on'' and ``off'' states of a 
molecular switch and provide a proof of principle that both 
forms can be reversibly interconverted using an external 
electrostatic field.

\section{Methods}
\label{sec:1}
To investigate the transport properties of the molecular junctions, we have
used a methodology which combines electronic structure calculations
employing density functional theory (DFT)
and Landauer transport theory.
The method has been described in detail previously \cite{Benesch2008}.
Here, we only provide the main aspects and some details specific to the
current application. 

In the electronic structure calculations, we have modelled 
the molecular junction using 
the bridge molecule and two clusters containing three layers 
of Au atoms (comprising a total of 88 Au 
atoms). The equilibrium structures of the 
keto and enol tautomers have been optimized using DFT methods. 
Specifically, as described 
elsewhere \cite{Benesch2006}, we have carried out a partial 
geometry optimization of a cluster consisting of the respective 
tautomer and the first layer of the Au atoms of each lead in 
order to determine realistic tautomer-lead binding geometries. 
Subsequently, two extra layers of Au atoms per lead were added 
to the optimized structures to obtain 
the models investigated in this work 
(see Fig. \ref{fig2}). Following previous work 
{\cite{Benesch2009}}, all calculations were carried out 
using the B3LYP hybrid correlation-exchange functional 
together with a SV(P) basis set \cite{Schafer1992}  
(Au atoms have been described using the ECP-60-MWB 
pseudopotential \cite{Andrae1990}) as implemented in 
TURBOMOLE \cite{Turbomole}. To describe the effect of infinite
leads we have added the surface self-energy  of a Au(111) surface to the 
atomic orbital energies of the Au atoms of the outer layer 
of both leads \cite{Benesch2008}.

The current for a given voltage was obtained from the 
Landauer formula \cite{Fisher1981}
\begin{equation} 
\label{eqlandauer}
I=\frac{2e}{h}\int T(E)\left[f_{L}(E)-f_{R}(E)\right]dE
\end{equation}
where  $f_{L}(E)$ and $f_{R}(E)$ are the Fermi 
distributions of the left and right lead, respectively \cite{Note1}, 
and $T(E)$ is the transmission function, which depends on the 
applied bias voltage. $T(E)$ contains the information about 
the junction and can be expressed as \cite{Meir1992}
\begin{equation}
\label{eqtrans}
T(E)= \mathbf{tr}_M [\Gamma_{L}(E)G_{M}^{\dagger}(E)\Gamma_{R}(E)G_{M}(E)]
\end{equation}
Here, $G_M(E)$ denotes the molecular Green's function and the 
functions $\Gamma_{L/R}$ describe 
broadening of molecular states due to coupling  with the metallic leads.
The methodology to obtain these functions is described in 
Ref.\ \cite{Benesch2008}.

\section{Results and Discussion}
\label{sec:results}

\begin{figure}[t]
\centering
\includegraphics[width=0.70\textwidth]{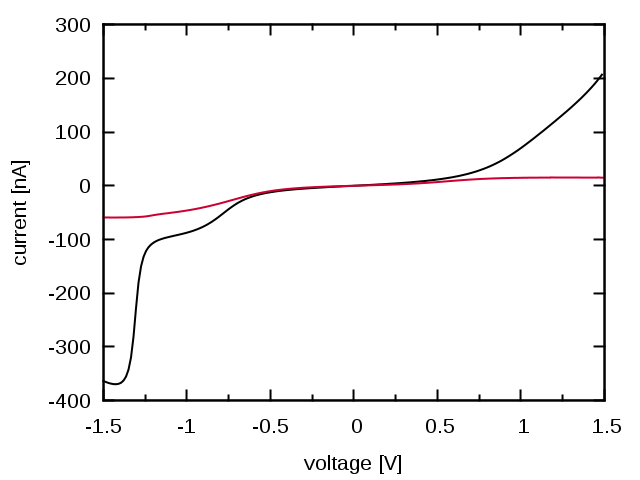}
\caption{\label{fig3} Current-voltage characteristics of the two tautomeric 
forms of the molecular junction depicted in Fig. \ref{fig2}. 
The black and the red lines depict the current of the enol and 
keto tautomers, respectively.}     
\end{figure}

Fig.\ \ref{fig3} shows the current-voltage characteristics
calculated for  T2HP and T2PY (see Fig. \ref{fig2}). 
The results reveal very different transport properties of the two
tautomers. While the keto form exhibits a rather low current, the enol form
facilitates a significantly larger current for bias voltages $|V| > 0.5$ V. 
Thus, for a given bias voltage, the two tautomers realize different
conductance states of the molecular junction and can be associated with 
``on'' (enol) and ``off'' (keto) states of a molecular nano switch.

\begin{figure}[t]
\centering
\includegraphics[width=0.70\textwidth]{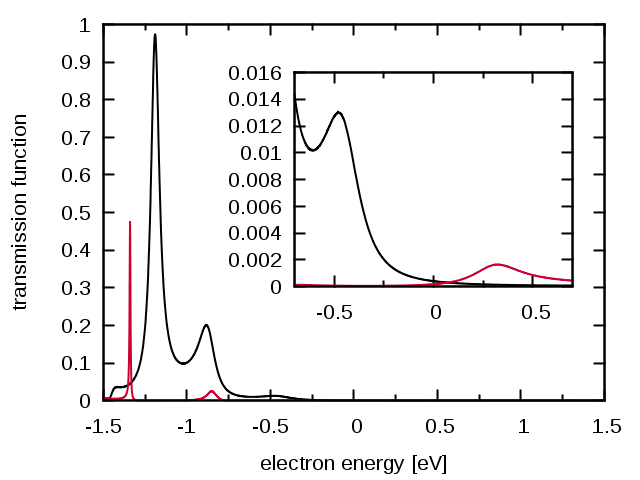}
\caption{\label{fig4}Transmission as a function of electron energy 
(relative to the Fermi level) at zero bias voltage. The black 
and red lines correspond to the enol and keto forms, respectively. 
The inset depicts the region close to the Fermi level.}
\end{figure}

\begin{figure*}[!t]
  \centering
  \subfigure{\includegraphics[width=0.47\textwidth]{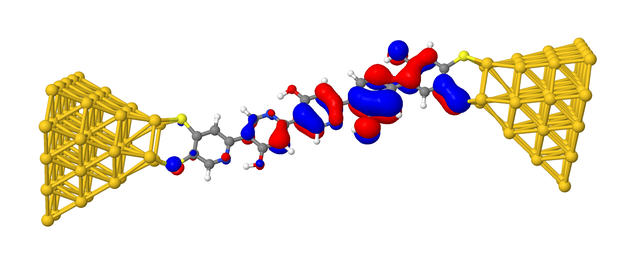} \label{figenol140}}  \subfigure
  {\includegraphics[width=0.47\textwidth]{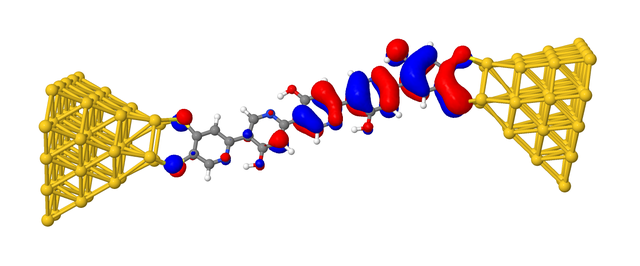} \label{figketo149}} \\ 
  \quad \subfigure{\includegraphics[width=0.39\textwidth]{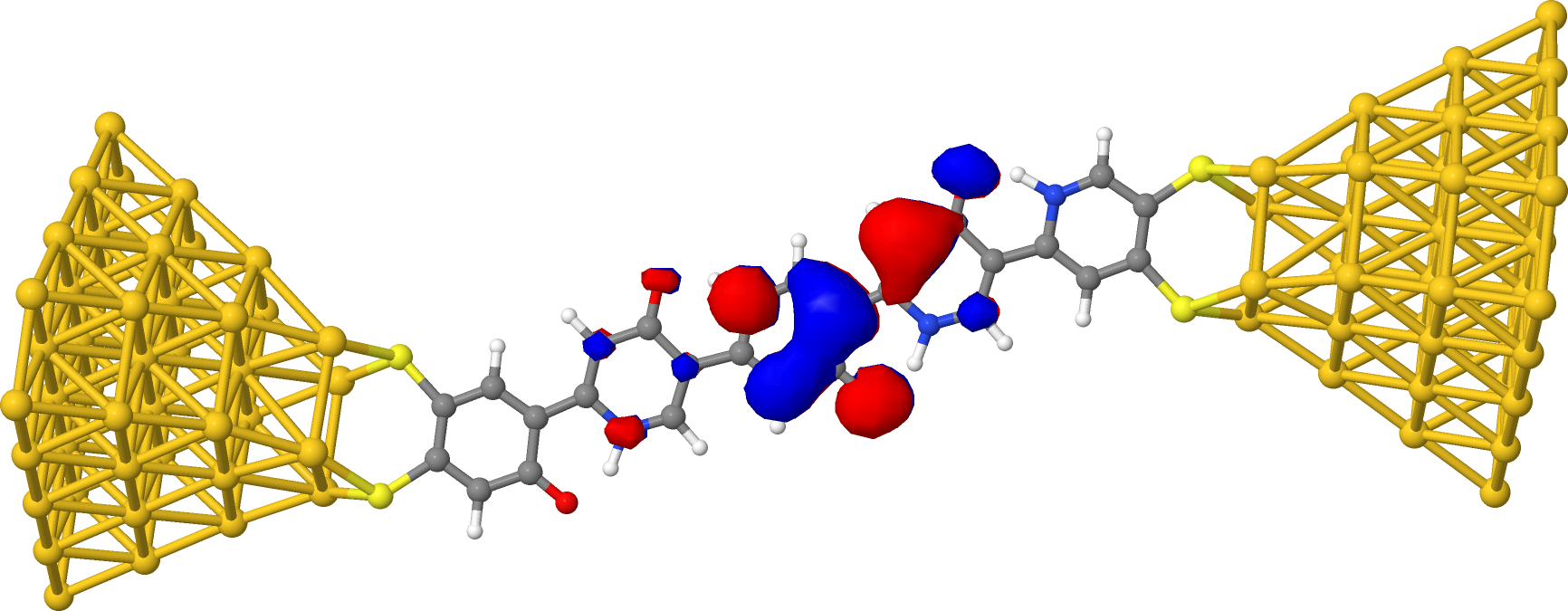} \label{figenol141}} \\  
  \caption{\label{fig5} Molecular orbitals associated to the most significant peaks in $T(E)$ for the enol (top) and keto (bottom) tautomers (see text for details).} 
\end{figure*}

To obtain insight into the mechanism underlying the very different conductance
properties of the two tautomers, we have analyzed their transmission functions
 $T(E)$ (see eq.\ (2)), depicted for zero bias 
voltage in Fig.\ \ref{fig4}.
The current is determined by the values of the transmission function 
close to the Fermi
energy. Both tautomers exhibit low values of transmission in the immediate
vicinity of the Fermi energy resulting in low current for 
bias voltages $|V| < 0.5$ V. Significant values of transmission are observed
for energies $ -1.5$ eV $ < E < -0.5$ eV, below the Fermi energy. 
The structures of the transmission function in this energy range 
reveal a pronounced
difference between the two tautomers. In particular, the transmission peaks of
the enol form are significantly broader, indicating a stronger coupling of the
molecular bridge to the leads and, thus, resulting in the larger current
observed for bias voltages $|V| > 0.5$ V (cf.\ Fig.\ \ref{fig3}).
This finding is corroborated by an analysis of the molecular orbitals 
associated to
the most relevant structures in the transmission function, depicted in 
Fig.\ \ref{fig5}.  In the case of 
the enol form, the peak at $E=-1.25$ eV is associated to the 
$\pi$-like orbitals (a) and (b) that extend over the molecule. 
On the other hand, 
the orbital (c) in Fig. \ref{fig5}. which is responsible for the peak 
at $E=-1.4$ eV in the keto 
tautomer, is more localized and 
has most of its density in the center of the molecule. The 
difference in the extension of the orbitals of the enol and keto 
forms can be explained in terms of the degree of conjugation 
present in both systems. Whereas the enol form is mainly characterized 
by a delocalized valence bond structure, the contribution of this 
class of valence bond structures to the keto tautomer can be 
expected to be less important.

\begin{figure}
\centering
\includegraphics[width=0.70\textwidth]{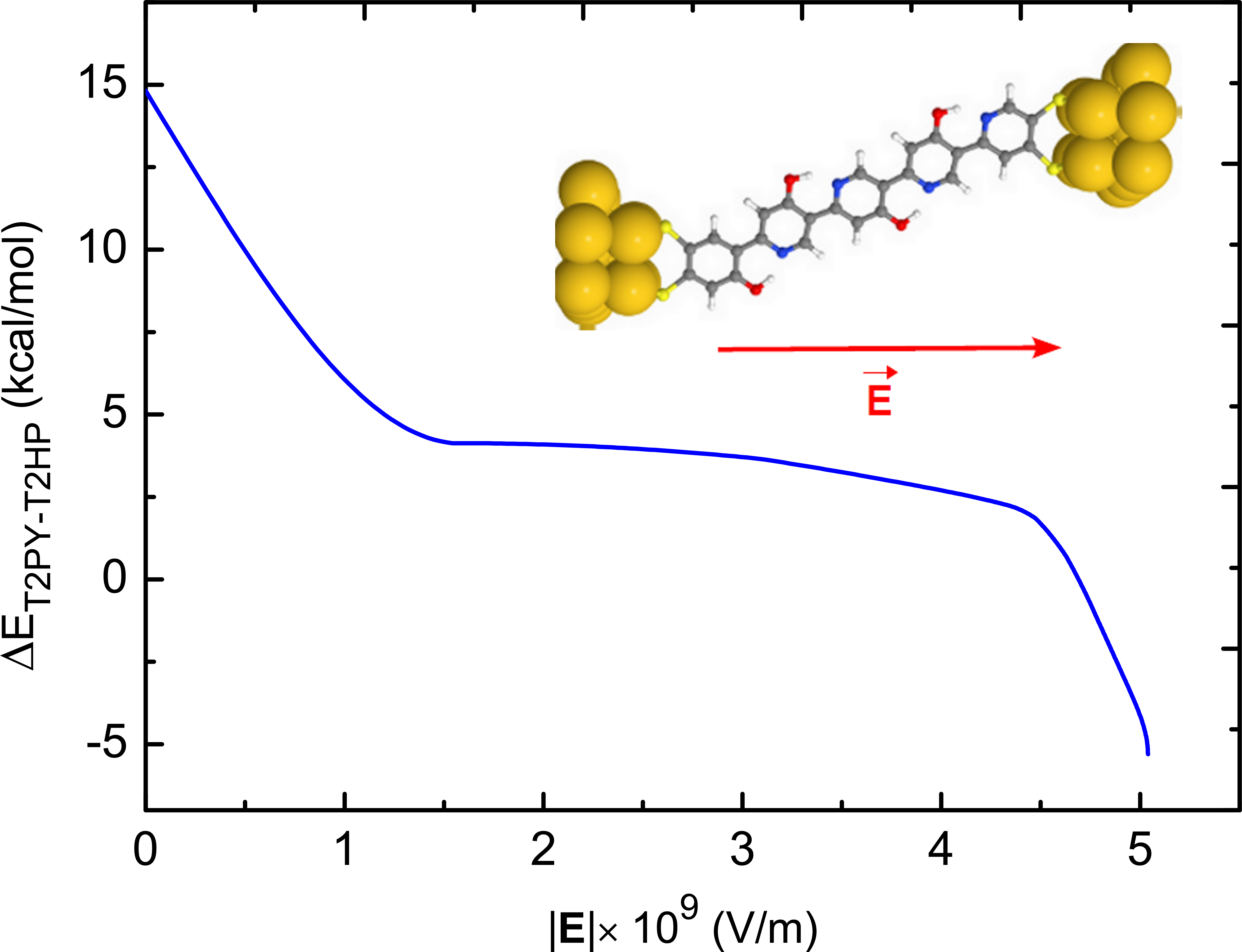}
\caption{\label{fig6} Energy difference between T2PY and T2HP for different values of the applied external electrostatic field. The inset shows the direction of the field.}
\end{figure}

The results presented above suggest that T2PH and T2PY may be used as 
the ``on'' and ``off'' states in a molecular nano switch. To achieve
this functionality, a mechanism for a reversible transformation between 
the two tautomers,
i.e. a proton transfer, is required. For this purpose, the different charge 
distributions of the two tautomers may be exploited. 
As is well known, proton transfer reactions
may be triggered by an external electrical field \cite{Matta2011}. To investigate this possibility, 
we have studied the relative stability of both tautomers in 
the presence of a constant external electrostatic field oriented 
in a direction along the axis of the junction 
(see Fig. \ref{fig6}). As a measure of the relative 
stability of the keto and enol forms, Fig. \ref{fig6} depicts the
difference of their ground state energies  as a function of the strength
 of the applied external field. The results show that without external
 field,  the most stable tautomer is the enol form. 
This is due to the stabilization obtained by aromaticity and 
conjugation. Application of an external electric field modifies 
the situation. Upon a gradual increase in the strength of the field the energy 
difference between the enol and the keto forms decreases and for 
values larger than $4.73\times10^{9}$ $\nicefrac{\text{V}}{\text{m}}$ 
an inversion of the relative stabilities of the tautomers can be 
obtained. This provides a proof of principle that a reversible 
interconversion between the T2HP and T2PY using an external electrostatic field 
is possible. In a more general context, the possibility to control the 
conductance by an external
electrostatic field, may also be used  as a 
mechanism for a single 
molecule transistor.

We finally mention some possible extensions of the present study.
First, for the specific system investigated, the largest difference in
current occurs only for larger voltages. This is due to the fact that the
most relevant molecular orbitals have energies  further away from the Fermi
energy. An optimization of the functionality can thus be achieved 
by using chemical substitutions that shift the
energies closer to the Fermi energy \cite{Pshe2013}.
Second, in the present study we have focused on the conductance properties of
the two tautomers at fixed nuclear geometry. An investigation of
the mechanism of switching requires the modelling of the 
potential energy surfaces along the reaction coordinates and the
nuclear dynamics under bias. Third, the external electrostatic field may also influence 
the conductance, e.g. by shifting the energy levels of the molecular bridge. Thus a more
complete description has to include the electric fields of the junction and the external electric field
in the transport calculations. This will be the subject of future work.

\section{Conclusions}    
We have explored the possibility to use a proton transfer reaction 
as a mechanism for switching a molecular junction between different
conductance states.
As a specific example, we have analyzed the conductance 
properties of T2HP and T2PY enol-keto tautomers. 
We have shown that both forms exhibit very different electrical currents, 
which allows their use as the ``on'' and ``off'' states of a
molecular nano switch. 
The origin of this difference has been explained in terms of the 
different electronic structures characterizing both tautomers. 
Furthermore, we have demonstrated that switching between the two conductance state
may be achieved by an external electrostatic field which triggers 
a proton transfer reaction that interconverts the two tautomers reversibly. 
The results of this theoretical study suggest that
keto-enol tautomers can be used as building blocks of a 
nanoscale molecular switch or, more generally, a molecular transistor.

\acknowledgement
This work has been supported by the the German-Israeli Foundation for Scientific Development (GIF), the Deutsche Forschungsgemeinschaft 
(DFG) through the Cluster of Excellence ``Engineering  
of Advanced Materials'' (EAM), SFB 953 and a research grant,  
as well as 
projects, CTQ2012-36966 (MICINN), and UAH2011/EXP-041 (UAH). 
ALS acknowledges the research grant 
of the National Science Centre of Poland 2011/01/M/ST2/00561.
Generous allocation of computing time at the computing centers in 
Erlangen (RRZE), 
Munich (LRZ), and J{\"u}lich (JSC)  is greatly acknowledged.


\providecommand{\latin}[1]{#1}
\providecommand*\mcitethebibliography{\thebibliography}
\csname @ifundefined\endcsname{endmcitethebibliography}
  {\let\endmcitethebibliography\endthebibliography}{}

\end{document}